\documentclass[aps,pre,amsmath,amssymb,twocolumn]{revtex4}
\usepackage{graphicx}
\usepackage{epstopdf}
\usepackage{pstricks}
\usepackage{amsmath}
\usepackage{xcolor}
\usepackage{subcaption}
\newcommand{\be}{\begin{equation}}
\newcommand{\ee}{\end{equation}}
\newcommand{\ba}{\begin{array}{l}}
\newcommand{\ea}{\end{array}}
\newcommand{\re}[1]{(\ref{#1})}
\newcommand{\ci}[1]{\cite{#1}}
\newcommand{\banonum}{\begin{eqnarray*}}
\newcommand{\eanonum}{\end{eqnarray*}}
\newcommand{\baa}{\begin{eqnarray}}
\newcommand{\eaa}{\end{eqnarray}}
\newcommand{\bfr}{\begin{flushright}}
\newcommand{\efr}{\end{flushright}}
\newcommand{\bfl}{\begin{flushleft}}
\newcommand{\efl}{\end{flushleft}}
\newcommand{\lab}[1]{\label{#1}}

\begin{document}

\title{Discrete Nonlocal Nonlinear Schr\"{o}dinger equation on Metric Graphs: Dynamics of PT-Symmetric Solitons in Discrete Networks}

\author{M. Akramov$^a$, F.~Khashimova$^{b}$, D. Matrasulov$^{c}$}

\affiliation{
$^a$National University of Uzbekistan, Vuzgorodok, Tashkent 100174, Uzbekistan\\
${^c}$Turin Polytechnic University in Tashkent, 17 Niyazov Str.,
100095,  Tashkent, Uzbekistan\\
$^b$ Navoiy State Mining Institute, 27 Janubiy str., Navoiy, Uzbekistan}

\begin{abstract}
 We consider PT-symmetric, discrete nonlocal nonlinear Schr\"{o}dinger equation on metric
 graphs.  Soliton solutions are obtained for simplest graph topologies, such as star and tree graphs.  Integrability of the problem is shown by proving existence of infinite number of conservation laws. 

\end{abstract}
\maketitle

\section{Introduction}
PT-symmetric nonlocal nonlinear Schr\"{o}dinger (NNLS) equation was introduced first by  Ablowitz and
Musslimani in \ci{AM2013}. Remarkable feature of NNLS equation is its integrability and approving soliton solutions which can be obtained using the inverse scattering method.  Different aspects of nonlinear nonlocal Schr\"{o}dinger equation, such as integrability, various soliton solutions and their properties have been studied during past few years \ci{AM2013}-\ci{Panos2020}. In \ci{AM2014} discrete  version of NNLS equation has been considered and its integrability was shown. Later, in \ci{AM2014}, discrete version of NNLS equation was introduced. In particular, it was shown, in analogy with the Ablowitz-Laddik equation, that it also integrable and approved soliton solutions. Discrete nonlinear Schr\"{o}dinger equations describe wave/soliton propagation in optical waveguide arrays \ci{Array}, Toda lattices \ci{Toda}, Hirota's nonlinear network \ci{Hirota}, etc. In this paper we extend the study of the Ref.\ci{AM2014} to the case of branched 1D domains, modeling these latter in terms of metric graphs. Such graphs are defined as the set of wires connected to each other according to a rule, called the topology of a graph. Each wire(arm) is assumed to be assigned a length. In particular, we address the problem of PT-symmetric discrete nonlocal nonlinear Schr\"{o}dinger equation on metric graphs with the focus on the exact (soliton) solutions and soliton dynamics. Motivation for the study soliton dynamics networks and branched structures comes from the fact that the wave dynamics in these latter is richer and more tunable. One can, e.g. achieve needed wave transport regime by tuning the network architecture or reflection transmission and the nodes (vertices). Another important issue related to modeling of soliton dynamics in networks is caused by the fact that in practical applications (e.g., in optics and optoelectronics) for signal transfer one often uses optical fiber networks. Therefore studying the soliton propagation in optical networks is of importance from the viewpoint of controlling the signal transfer, reducing the signal losses and optoelectronic device optimization. We note that earlier the soliton dynamics in networks described in terms of nonlinear
Schr\"{o}dinger \ci{}, Manakov \ci{},sine-Gordon \ci{} and nonlinear Dirac \ci{} equations has attracted much attention. The model we propose describes soliton propagation in discrete optical waveguide arrays, Toda lattices and Hirota's nonlinear network with self-induced PT-symmetric nonlinearity. This paper is organized as follows. In the next section, we briefly recall
discrete NNLS equation on a line, following the Ref.\cite{AM2014}. Section III presents formulation and solution of the problem, including both numerical and analytical solutions, for a star branched network. In section IV we extend the study for tree-like networks. Finally, section V presents some concluding remarks.

\section{Discrete Nonlocal Nonlinear Schr\"{o}dinger equation on a line}

Discrete version of the Ablowitz-Musslimani nonlocal nonlinear Schr\"{o}dinger equation  can be written as \ci{AM2014}
\begin{eqnarray}
i\frac{dQ_n}{dt}=Q_{n+1}-2Q_n+Q_{n-1}+Q_n Q_{-n}^* (Q_{n+1}+Q_{n-1}). \label{eq1}
\end{eqnarray}
where $n$ being integer in $(-\infty,+\infty)$. This equation can be considered as PT-symmetric, nonlocal analog of the classical discrete nonlinear Schr\"{o}dinger  equation \ci{AL1,AL2}, i.e. new and simple reduction of the well-known Ablowitz-Ladik scattering problem. Integrability and soliton solutions of \re{eq1} was studied in \ci{AM2014}. 
In particular, single soliton solution of Eq.\eqref{eq1} can be obtained using the inverse scattering method and is given by
\begin{eqnarray}
Q_n(t)=-\frac{(z_1 \Bar{z}_1)^{-1} (z_1^2-\Bar{z}_1^2) e^{i\Bar{\varphi}_1} e^{-2i\Bar{\omega}_1 t} \Bar{z}_1^{2n}       }{1+e^{ i(\varphi_1+\Bar{\varphi}_1) } e^{2i ( \omega_1-\Bar{\omega}_1 )t}   z_1^{-2n} \Bar{z}_1^{2n} }, \label{sol1}
\end{eqnarray}
where, $z_1>0$, $0<\Bar{z}_1<1$, $\omega_1=(z_1-z_1^{-1})/2$, $\Bar{\omega}_1=(\Bar{z}_1-\Bar{z}_1^{-1})/2$ and $\varphi_1$, $\Bar{\varphi}_1$ are arbitrary constants.
Let us introduce another one soliton solution by Ref.\cite{Zuo-Nong}
\begin{equation}
Q_n(t)=\frac{\varepsilon e^{\kappa n+\omega t}}{1+\varepsilon^2 \theta e^{(\kappa-\kappa^*)n+(\omega+\omega^*)t}},
\end{equation}
where $\omega=-2i(\cosh\kappa-1), \; \theta=\frac{1}{4}\sinh^{-2}\frac{\kappa-\kappa^*}{2}$ and $\varepsilon, \; \kappa$ are constant numbers.
In the Ref.\cite{Zhang} obtained one soliton solution by bilinearisation-reduction approach in the form
\begin{equation}
Q_n(t)=\frac{e^{2k_1}-e^{-2k_1^*}}{e^{k_1+k_1^*}(e^{-2k_1 n+2i\xi_1 t}-e^{-2k_1^* n+2i\xi_1^* t})},
\end{equation}
where $\xi_1=(e^{2k_1}-2+e^{-2k_1})/2$ and $k_1$ is a constant number.

Existence of infinitely many conservative quantities was also shown in \ci{AM2014}. The first few of them can be written as
\begin{gather}
C_1=-\sum_{n=-\infty}^{+\infty} Q_n Q_{1-n}^*, \label{C1} \\
C_2=-\sum_{n=-\infty}^{+\infty} \left[  Q_n Q^*_{2-n}+\frac{1}{2}(Q_n Q^*_{1-n})^2  \right], \label{C2} \\
C_3=\prod_{n=-\infty}^{+\infty} (1+Q_n Q^*_{-n}). \label{C3}
\end{gather}
Eq.\re{eq1} approves also a Hamiltonian as conserving quantity and non canonical brackets which are given by
\begin{gather}
H=\sum_{n=-\infty}^{+\infty} \left[   Q^*_{-n}(Q_{n+1}+Q_{n-1})-2\log(1+Q_n Q^*_{-n})  \right], \label{H}\\
\{  Q_m, Q^*_{-n}  \}=-i(1+Q_n Q^*_{-n})\delta_{n,m},\\
\{  Q_n, Q_m  \}=\{  Q^*_{-n}, Q^*_{-m}  \}=0.
\end{gather}
By using last equations Eq.\eqref{eq1} can be written in the form
\begin{eqnarray}
i\frac{dQ_n}{dt}=\{ Q_n, H \}.
\end{eqnarray}

In the next section we use some of these conserving quantities to derive vertex boundary conditions for DNNLS equation on metric graphs and solution \re{sol1} to obtain its soliton solutions.

\section{PT-symmetric nonlocal solitons in discrete star shaped network}
\begin{figure}[t!]
\includegraphics[width=80mm]{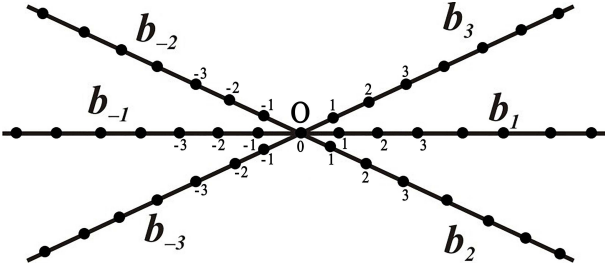}
\caption{Star graph with six bonds.} \label{pic1}
\end{figure}
The  DNNLS equation on a line considered in the previous section can be extended to the case of branched quasi-one dimensional domains called metric graphs. Simplest metric graph, where such a problem can be considered has a six bonds connected in a single node and called metric star graph (see, Fig.1). It represents a branched 1D lattice with six arms. The individual lattice sites in the graph are denoted as $(\pm j,n)$, where $\pm j=\pm 1,\pm 2,\pm 3$ are the bond's number and $n$ corresponds to a lattice site on each bond.
For the left handed $(-j=-1,-2,-3)$ bonds $n \in b_{-j}=\{0,-1,-2,...\}$, where $(-j,0)$ means the branching point.
For the right handed $(j=1,2,3)$ bonds $n \in b_j=\{1,2,3,...\}$, where $(j,1)$ stand for the points nearest to the vertex.

Consider the following discrete version of PT-symmetric, nonlocal nonlinear Schr\"odinger equation which is written on the each bond of the  graph in Fig. 1 as
\begin{gather}
i\frac{dQ_{\pm j,n}}{dt}=Q_{\pm j,n+1}-2Q_{\pm j,n}+Q_{\pm j,n-1}+\nonumber\\
\sqrt{\beta_j \beta_{-j}}Q_{\pm j,n} Q_{\mp j,-n}^* (Q_{\pm j,n+1}+Q_{\pm j,n-1}), \label{eq2}
\end{gather}
where $\beta_{\pm j}$ are the nonlinearity coefficients and $(\pm j,n) \notin \{(-j,0),(j,1)\}$.
Continuum version of the NNLS equation on such domain was studied in detail recently in \ci{Mashrab2022}. To solve Eq.\re{eq2}, one needs to impose boundary conditions at the branching point, $O$. In case of continuum NNLSE such boundary conditions can be derived from energy and norm conservation laws \ci{Mashrab2022}. However, for  discrete evolution equations on graphs derivation of vertex boundary conditions is complicated. Therefore, following the Ref. \ci{zar2011}, we explore first the dynamics of solitons in the vicinity of the vertex, i.e. in the sites $(-j,1),(j,0)$.   Using the same approach as that for Ablowitz-Laddik equation on metric graphs, considered in \ci{zar2011}, one can show that at the virtual sites, $(-j,1),(j,0)$,
Eq.\eqref{eq2} can be obtained from the following equation of motion
\begin{eqnarray}
i\frac{dQ_{\pm j, n}}{dt}=\{  Q_{\pm j,n}, H  \}, \label{eq3}
\end{eqnarray}
at $(\pm j,n) \notin \{(-j,0),(j,1)\}$. Here the non canonical Poisson brackets are determined as \ci{zar2011}
\begin{gather}
\{  Q_{\pm j, m}, Q_{\mp j', -n}  \} = -i(1+\sqrt{\beta_j \beta_{-j}} Q_{\pm j,n} Q_{\mp j,-n}^*)\delta_{jj'} \delta_{nm},\nonumber\\
\{  Q_{\pm j, n}, Q_{\pm j', m}  \} = \{  Q_{\pm j, -n}^*, Q_{\pm j', -m}^*  \} = 0. 
\end{gather}
and $H$ is the Hamiltonian for DNNLS equation given by
\begin{widetext}
\begin{eqnarray}
H=\sum_{j=1}^3 \bigg[
\sum_{n=0}^{-\infty} \left(   Q_{j,-n}^*(Q_{-j,n+1}+Q_{-j,n-1})-\frac{2}{\sqrt{\beta_j \beta_{-j}}}\log(1+\sqrt{\beta_j \beta_{-j}}Q_{-j,n} Q_{j,-n}^*)   \right)+\nonumber\\
\sum_{n=1}^{+\infty} \left(   Q_{-j,-n}^*(Q_{j,n+1}+Q_{j,n-1})-\frac{2}{\sqrt{\beta_j \beta_{-j}}}\log(1+\sqrt{\beta_j \beta_{-j}} Q_{j,n}Q_{-j,-n}^*)   \right)
\bigg], \label{hamiltonian}
\end{eqnarray}
\end{widetext}
where at the sites, $(-j,1),(j,0)$ we assumed the following relations:
\begin{gather}
Q_{-j, 1}=\alpha^{(-j)}_1 Q_{1,1}+\alpha^{(-j)}_2 Q_{2,1}+\alpha^{(-j)}_3 Q_{3,1},\nonumber\\
Q_{j, 0}=\alpha^{(j)}_{-1} Q_{-1,0}+\alpha^{(j)}_{-2} Q_{-2,0}+\alpha^{(j)}_{-3} Q_{-3,0}  \label{assumption}
\end{gather}
with appropriate coefficients, $\alpha^{(\pm j)}_{\pm j}$.

Explicitly,  Eq.\eqref{eq3} at the sites $(-j,0), (j,1)$ can be written, respectively as
\begin{widetext}
\begin{gather}
i\frac{dQ_{-j,0}}{dt}=\alpha^{(-j)}_1 Q_{1,1}+\alpha^{(-j)}_2 Q_{2,1}+\alpha^{(-j)}_3 Q_{3,1}-2Q_{-j,0}+Q_{-j,-1}+\nonumber\\
\sqrt{\beta_j \beta_{-j}}Q_{-j,0} \left(\alpha^{(j)}_{-1} Q^*_{-1,0}+\alpha^{(j)}_{-2} Q^*_{-2,0}+\alpha^{(j)}_{-3} Q^*_{-3,0}\right)
\left( \alpha^{(-j)}_1 Q_{1,1}+\alpha^{(-j)}_2 Q_{2,1}+\alpha^{(-j)}_3 Q_{3,1}+Q_{-j,-1} \right), \label{eq4}\\
i\frac{dQ_{j,1}}{dt}= Q_{j,2}-2Q_{j,1}+ \alpha^{(j)}_{-1} Q_{-1,0}+\alpha^{(j)}_{-2} Q_{-2,0}+\alpha^{(j)}_{-3} Q_{-3,0}+\nonumber\\
\sqrt{\beta_j \beta_{-j}}Q_{j,1} Q_{-j,-1}^* \left( Q_{j,2}+\alpha^{(j)}_{-1} Q_{-1,0}+\alpha^{(j)}_{-2} Q_{-2,0}+\alpha^{(j)}_{-3} Q_{-3,0} \right). \label{eq5}
\end{gather}
\end{widetext}

Furthermore, we suppose that the soliton solution of DNNLS equation \re{eq2} on a star graph presented in Fig.1 is given by
\begin{gather}
Q_{\pm j,n}(t)=\frac{1}{\sqrt{\beta_{\pm j}}} Q_{n}(t), \label{solution}
\end{gather}
where $Q_{n}(t)$ is the soliton solution of the DNNLS equation on a discrete line, (i.e. on an  infinite chain)  given by Eq.\eqref{eq1}.

Comparing Eqs.\eqref{eq1},\eqref{eq4},\eqref{eq5} and using Eq.\eqref{solution}, one can find two relations between the coefficients  $\alpha^{(\pm j)}_{\pm j}$ and $\beta^{(\pm j)}_{\pm j}$:

The first choice is given by
\begin{eqnarray}
\alpha^{(\pm j)}_{\mp k}=\frac{1}{3}\sqrt{\frac{\beta_{\mp k}}{\beta_\pm j}}, \quad k=1,2,3,
\end{eqnarray}
and 
\begin{eqnarray}
\alpha^{(\pm j)}_{\mp k}=\frac{1}{\sqrt{\beta_{\pm j}\beta_{\mp k}} \left(   \frac{1}{\beta_{\pm 1}}+\frac{1}{\beta_{\pm 2}}+\frac{1}{\beta_{\pm 3}}  \right)}. \label{alpha}
\end{eqnarray}
Substituting Eq.\eqref{alpha} into Eq.\eqref{assumption}, one can obtain the following "quasi-boundary" conditions at the vertex:
\begin{gather}
\sqrt{\beta_{-1}}Q_{-1,1}=\sqrt{\beta_{-2}}Q_{-2,1}=\sqrt{\beta_{-3}}Q_{-3,1},\nonumber\\
\frac{Q_{-1,1}}{\sqrt{\beta_{-1}}}+\frac{Q_{-2,1}}{\sqrt{\beta_{-2}}}+\frac{Q_{-3,1}}{\sqrt{\beta_{-3}}}=
\frac{Q_{1,1}}{\sqrt{\beta_{1}}}+\frac{Q_{2,1}}{\sqrt{\beta_{2}}}+\frac{Q_{3,1}}{\sqrt{\beta_{3}}}, \label{vbc1}
\end{gather}
and
\begin{gather}
\sqrt{\beta_{1}}Q_{1,0}=\sqrt{\beta_{2}}Q_{2,0}=\sqrt{\beta_{3}}Q_{3,0},\nonumber\\
\frac{Q_{1,0}}{\sqrt{\beta_{1}}}+\frac{Q_{2,0}}{\sqrt{\beta_{2}}}+\frac{Q_{3,0}}{\sqrt{\beta_{3}}}=
\frac{Q_{-1,0}}{\sqrt{\beta_{-1}}}+\frac{Q_{-2,0}}{\sqrt{\beta_{-2}}}+\frac{Q_{-3,0}}{\sqrt{\beta_{-3}}}. \label{vbc2}
\end{gather}

These quasiboundary conditions are fulfilled by the solutions \re{solution}, provided the nonlinearity coefficients, $\beta_{\pm j}$ fulfill the constraints given by
\begin{eqnarray}
\frac{1}{\beta_{-1}}+\frac{1}{\beta_{-2}}+\frac{1}{\beta_{-3}}=\frac{1}{\beta_{1}}+\frac{1}{\beta_{2}}+\frac{1}{\beta_{3}}.\label{constraint}
\end{eqnarray}

\begin{figure}[t!]
\includegraphics[width=90mm]{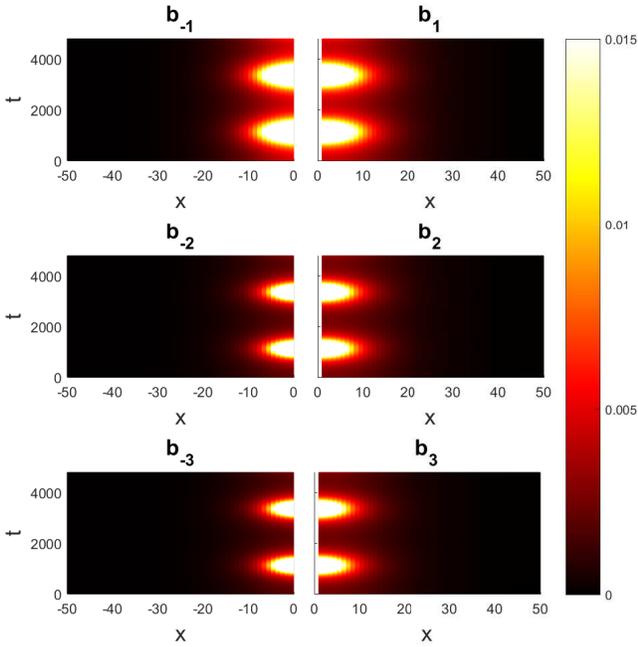}
\caption{Contour of the (breathing soliton) solution of DNNLS equation\eqref{eq2} on metric star graph given by Eq.\re{solution}. The nonlinearity coefficients are chosen as $\beta_{-1}=1, \; \beta_{1}=1.15, \; \beta_{-2}=2.19, \; \beta_{2}=1.92, \; \beta_{-3}=2.42, \; \beta_{3}=2.09$}. \label{fig:pic2}
\end{figure}

\subsection{Numerical results}

The solution given by Eq. \re{solution} presents a breathing soliton, which is valid for the case, when the constraints given by Eq.\re{constraint} is fulfilled. In general case, i.e., when the constraint is not fulfilled, one needs to solve Eq.\re{eq2} numerically.  In Fig.2 contour plot of the (breathing) soliton solution of Eq.\re{eq2} for the quasiboundary conditions given by Eqs.\eqref{vbc1} and \eqref{vbc2} is presented. The plot is obtained using the analytical formula \re{solution} for the values of parameters  $\beta_{-1}=1, \; \beta_{1}=1.15, \; \beta_{-2}=2.19, \; \beta_{2}=1.92, \; \beta_{-3}=2.42, \; \beta_{3}=2.09$, which fulfill the sum rule \re{constraint}. 
Fig. 3 presents numerically obtained travelling wave solution of DNNLS equation on graph, \re{eq2} for the case, when the sum rule given by Eq.\re{constraint} is fulfilled. Reflectionless transmission of soliton through the vertex can be observed here, unlike the Fig.4, where numerically obtained travelling wave solutions are plotted for the values of $\beta_{\pm j}$, which do not fulfill constraint \re{constraint}. Reflection of soliton at the branching area can be clearly seen from these plots.
In numerical solution of Eq.\re{eq2} the initial conditions are chosen on the $b_{-1}$ and $b_1$ bonds, in the form of Gaussian wave packet given by
\begin{eqnarray}
Q_{\pm 1,n}=\frac{A}{\sqrt{\beta_{\pm 1} \sigma \sqrt{2\pi}}} \exp\left(\pm i k_0 n-\bigg(\frac{n-n_0}{2\sigma^2}\bigg)^2\right),
\end{eqnarray}
where $A,\; k_0, \; \sigma$ and $n_0$ are constants.

\begin{figure}[t!]
\includegraphics[width=90mm]{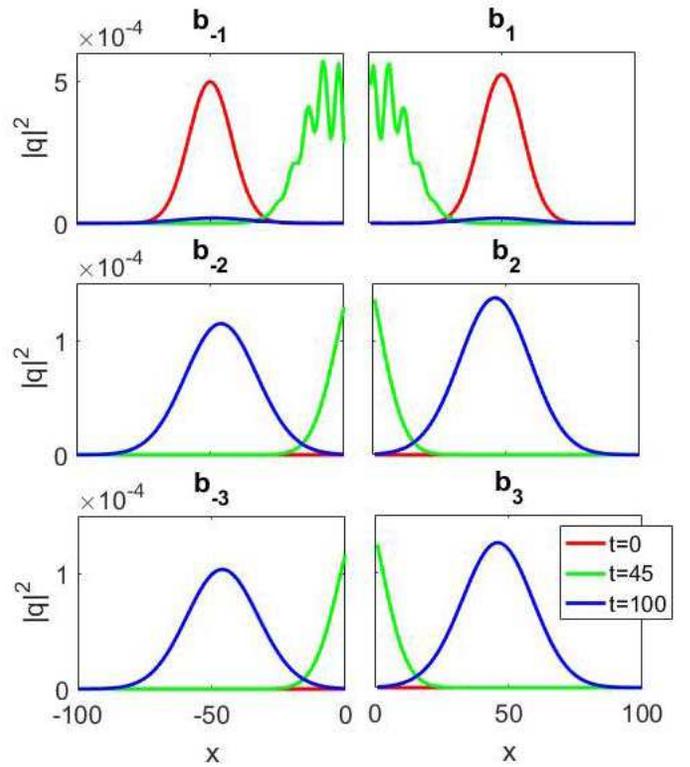}
\caption{Numerically obtained travelling wave solution of DNNLS equation on the star graph when the sum rule, \re{constraint} is fulfilled. The nonlinearity coefficients are chosen as $\beta_{-1}=1$, $\beta_1=1.15$, $\beta_{-2}=2.19$, $\beta_2=1.91$, $\beta_{-3}=2.42$, $\beta_3=2.09$. The initial conditions are given on the bonds $b_{-1}$ and $b_1$.} \label{fig:fulfilled}
\end{figure}

\section{Conservation laws and integrability}
In \ci{AM2014}, integrability of DNNLS equation on a line was shown by proving existence of infinitely many conservation laws. Similar thing is possible for DNNLSE equation on graph, too. For Eq.\re{eq2}, two fundamental  conservative quantities, the norm, $C_1$ and the energy, $E$  (for star graph, in Fig.1) can be written as, respecively 
\begin{eqnarray}
C_1=-\sum_{j=1}^3 \left[   \sum_{n=0}^{-\infty} Q_{-j,n} Q_{j,1-n}^* + \sum_{n=1}^{+\infty} Q_{j,n} Q_{-j,1-n}^*  \right], \label{eq6}
\end{eqnarray}
 and
\begin{widetext}
\begin{gather}
E=\sum_{j=1}^3 \bigg[
\sum_{n=-1}^{-\infty} \left(   Q_{j,-n}^*(Q_{-j,n+1}+Q_{-j,n-1})-\frac{2}{\sqrt{\beta_j \beta_{-j}}}\log(1+\sqrt{\beta_j \beta_{-j}}Q_{-j,n} Q_{j,-n}^*)   \right)+\nonumber\\
\sum_{n=2}^{+\infty} \left(   Q_{-j,-n}^*(Q_{j,n+1}+Q_{j,n-1})-\frac{2}{\sqrt{\beta_j \beta_{-j}}}\log(1+\sqrt{\beta_j \beta_{-j}} Q_{j,n}Q_{-j,-n}^*)   \right) +\nonumber\\
(\alpha^{(j)}_{-1} Q^*_{-1,0}+\alpha^{(j)}_{-2} Q^*_{-2,0}+\alpha^{(j)}_{-3} Q^*_{-3,0})(\alpha^{(-j)}_1 Q_{1,1}+\alpha^{(-j)}_2 Q_{2,1}+\alpha^{(-j)}_3 Q_{3,1}+Q_{-j,-1})-\nonumber\\
\frac{2}{\sqrt{\beta_j \beta_{-j}}}\log(1+\sqrt{\beta_j \beta_{-j}}Q_{-j,0} (\alpha^{(j)}_{-1} Q^*_{-1,0}+\alpha^{(j)}_{-2} Q^*_{-2,0}+\alpha^{(j)}_{-3} Q^*_{-3,0}))+\nonumber\\
Q_{-j,-1}^*(Q_{j,2}+\alpha^{(j)}_{-1} Q_{-1,0}+\alpha^{(j)}_{-2} Q_{-2,0}+\alpha^{(j)}_{-3} Q_{-3,0})-\frac{2}{\sqrt{\beta_j \beta_{-j}}}\log(1+\sqrt{\beta_j \beta_{-j}} Q_{j,1}Q_{-j,-1}^*)
\bigg]. \label{energy}
\end{gather}
\end{widetext}

\begin{figure}[t!]
\includegraphics[width=90mm]{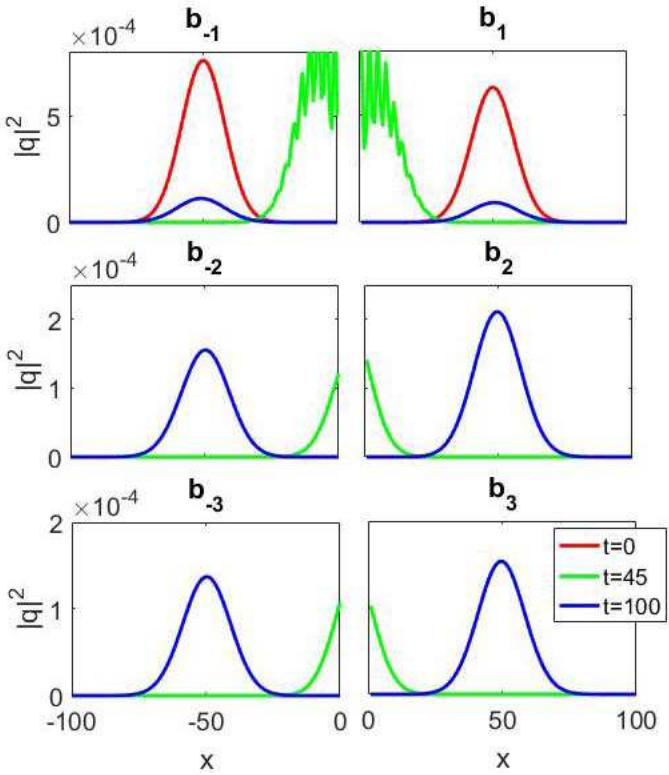}
\caption{Numerically obtained travelling wave solution of DNNLS equation on the star graph when the sum rule, \re{constraint} is broken. The nonlinearity coefficients are chosen as $\beta_{-1}=0.65$, $\beta_1=0.79$, $\beta_{-2}=2.7$, $\beta_2=2.09$, $\beta_{-3}=3.06$, $\beta_3=2.87$. The initial conditions are given on the bonds $b_{-1}$ and $b_1$.} \label{fig:broken}
\end{figure}

Conservation laws for these quantities are written in terms of their time derivatives as $\frac{d}{dt}C_1=0$ and   $\frac{d}{dt}E=0$.
Existence of infinitely many conserving quantities can be shown similarly to that for
DNNLS equation on a line studied in detail in \ci{AM2014}.

\section{Extension to a tree graph}

The above treatment of DNNLSE equation on a star branched network can be extended to the case of other branching topologies. Here we demonstrate this for a tree branched graph, presented in Fig.3. The graph consists of 10 semi-infinite and 4 finite bonds. These latter are determined as follows: $b_{-1},b_{-1mn}=\{-1,-2,-3,...\}$, $b_{-1m}=\{ 0,-1,-2,...,-N_m \}$, $b_{1},b_{1mn}=\{1,2,3,...\}$, $b_{1m}=\{ 1,2,3,...,N_m \}$, where $m,n=1,2$.

DNNLS equation on the each bond of tree graph can be written as
\begin{gather}
i\frac{dQ_{\pm e,n}}{dt}=Q_{\pm e,n+1}-2Q_{\pm e,n}+Q_{\pm e,n-1}+\nonumber\\
\sqrt{\beta_e \beta_{-e}}Q_{\pm e,n} Q_{\mp e,-n}^* (Q_{\pm e,n+1}+Q_{\pm e,n-1}), \label{eq15}
\end{gather}
where $e=\{ 1,1m,1mn \}$. We note that Eq.\eqref{eq15} is written at the sites, which do not touch the vertices (branching points), i.e., the sites in nearest vicinity of vertices. Similarly to that for the star graph, we impose the following conditions on four symmetric branching points:
\begin{gather}
Q_{-1mn,1}=\alpha_{-1m}^{(-1mn)} Q_{-1m,-N_m},\nonumber\\
Q_{-1m,-N_m+1}=\alpha_{-1m1}^{(-1m)} Q_{-1m1,0}+\alpha_{-1m2}^{(-1m)} Q_{-1m2,0},\nonumber\\
Q_{1mn,0}=\alpha_{1m}^{(1mn)} Q_{1m,N_m},\nonumber\\
Q_{1m,N_m+1}=\alpha_{1m1}^{(1m)} Q_{1m1,1}+\alpha_{1m2}^{(1m)} Q_{1m2,1}. \label{assumption1}
\end{gather}
Soliton solutions of DNNLS equation on each branch of the tree graph can be expressed in terms of the solution of DNNLS equation on a line as
\begin{gather}
Q_{\pm e,n}(t)=\frac{1}{\sqrt{\beta_{\pm e}}} Q_{n+s_{\pm e}}(t),
\lab{sol001}
\end{gather}
where $s_{\pm 1}=s_{\pm 1m}=n_0$, $s_{\pm 1mn}=n_0 \pm N_{m}$ and $n_0$ is the center of soliton.
Requiring fulfilling the "quasi-boundary" conditions \eqref{assumption1}, by the solution \re{sol001}, one obtains the following  relations between coefficients, $\alpha_{\pm e}^{(\mp e)}$ and nonlinearity parameters, $\beta_{\pm mn}$:
\begin{gather}
\alpha_{\pm 1mn}^{(\pm 1m)}=\sqrt{\frac{\beta_{\pm 1m}}{\beta_{\pm 1mn}}},\nonumber\\
\alpha_{\pm 1m}^{(\pm 1mn)}=\frac{1}{\sqrt{\beta_{\pm 1m} \beta_{\pm 1mn}}\left(  \frac{1}{\beta_{\pm 1m1}} + \frac{1}{\beta_{\pm 1m2}}  \right)}. \label{alpha1}
\end{gather}
\begin{figure}[t!]
\includegraphics[width=80mm]{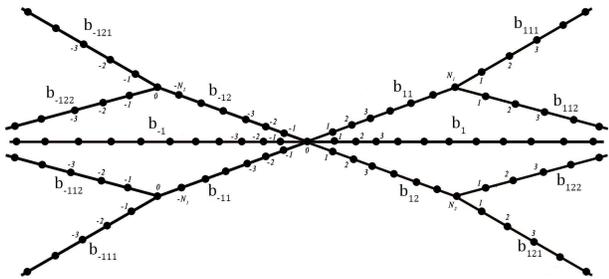}
\caption{Sketch of a tree graph}. \label{fig:tree}
\end{figure}
Substituting of Eq.\eqref{alpha1} into Eq.\eqref{assumption1} lead to the following "quasi-boundary conditions":
\begin{eqnarray}
\sqrt{\beta_{-1m1}} Q_{-1m1,1}=\sqrt{\beta_{-1m2}} Q_{-1m2,1}, \nonumber\\
\frac{Q_{-1m1,1}}{\sqrt{\beta_{-1m1}}} + \frac{Q_{-1m2,1}}{\sqrt{\beta_{-1m2}}} = \frac{Q_{-1m,-N_m}}{\sqrt{\beta_{-1m}}},\nonumber\\
\frac{Q_{-1m,-N_m+1}}{\sqrt{\beta_{-1m}}} = \frac{Q_{-1m1,0}}{\sqrt{\beta_{-1m1}}} + \frac{Q_{-1m2,0}}{\sqrt{\beta_{-1m2}}}, \label{tree_vbc1}
\end{eqnarray}
and
\begin{eqnarray}
\sqrt{\beta_{1m1}} Q_{1m1,0}=\sqrt{\beta_{1m2}} Q_{1m2,0}, \nonumber\\
\frac{Q_{1m1,0}}{\sqrt{\beta_{1m1}}} + \frac{Q_{1m2,0}}{\sqrt{\beta_{1m2}}} = \frac{Q_{1m,-N_m}}{\sqrt{\beta_{1m}}},\nonumber\\
\frac{Q_{1m,-N_m+1}}{\sqrt{\beta_{1m}}} = \frac{Q_{1m1,1}}{\sqrt{\beta_{1m1}}} + \frac{Q_{1m2,1}}{\sqrt{\beta_{1m2}}}. \label{tree_vbc2}
\end{eqnarray}
The above quasiboundary conditions fulfill by the solution Eq.\eqref{sol001}, if the following constraint in terms of $\beta_{\pm e}$ to be hold
\begin{eqnarray}
\frac{1}{\beta_{\pm 1m}}=\frac{1}{\beta_{\pm 1m1}}+\frac{1}{\beta_{\pm 1m2}}.
\lab{sumr}
\end{eqnarray}
The sum rule given by Eq.\re{sumr} ensures integrability of DNNLS equation \re{eq15} on tree graph domain presented in Fig.3 and approving  soliton solutions given by Eq.\re{sol001}. We note that the above utilized approach can be applied for solving DNNLS equation on discrete metric graphs of arbitrary topology, provided graph has two  for outgoing semi-infinite bonds.

\section{Conclusions}
In this paper, we studied dynamics of solitons described by PT-symmetric discrete nonlocal nonlinear Schr\"{o}dinger equation on networks by modeling these latter in terms of the discrete metric graphs. Integrability of the problem in case of fulfilling certain constraints given in terms of nonlinearity coefficients is shown. Exact soliton solutions which are valid for this case are obtained. For the case, when the constraints are broken, the problem is solved numerically. The model proposed in the paper can be used for describing soliton dynamics in discrete waveguide networks such as, e.g., branched Hirota lattices, discrete optical fiber arrays networks, where each branch has self-induced gain-loss.

\end{document}